\documentclass[%
reprint,
 amsmath,amssymb,
 aps,
 prd,
]{revtex4-2}

\usepackage{graphicx}
\usepackage{dcolumn}
\usepackage{bm}
\usepackage{xcolor}
\usepackage{float}
\usepackage{aas_macros}
\usepackage{adjustbox}

\newcommand{\LCDM}{\Lambda \textrm{CDM}}

\begin{document}
\preprint{APS/123-QED}

\title{Clustered unified dark sector cosmology: Background evolution and linear perturbations in light of observations}

\author{Mahmoud Hashim}
\author{Amr A. El-Zant}%
 \email{amr.elzant@bue.edu.eg}
\affiliation{
Centre for Theoretical Physics, The British University in Egypt, Sherouk City 11837, Cairo, Egypt
}

\date{\today}

\begin{abstract}
We consider unified dark sector models in which
the fluid can collapse and cluster into halos, allowing for hierarchical structure formation to proceed as in standard cosmology. We  show that both background evolution and linear perturbations tend towards those in $\LCDM$ 
as the clustered fraction $f \rightarrow 1$. 
We confront such models with various observational datasets, with emphasis on the relatively well motivated standard Chaplygin gas. We show that the strongest constraints come from secondary anisotropies  in the CMB spectrum,  
which prefer models with $f \rightarrow 1$. 
However, as a larger Hubble constant is allowed for smaller $f$,  
values  of $f \simeq 0.99$ (rather than tending to exact unity) 
are favored when late universe expansion 
data is included,  with $f \simeq 0.97$ and 
$H_0 \simeq 70 {\rm km/s/Mpc}$ allowed at the 2-$\sigma$ level.
{Such values of $f$ imply extremely efficient clustering into nonlinear 
structures. They may nevertheless be compatible with clustered fractions
in warm dark matter based cosmologies, which have similar minimal halo mass 
scales as the models considered here.}  
Tight CMB constraints on $f$ also apply to the 
generalized Chaplygin gas, except for models that are 
already quite close to $\LCDM$, in which case all values of $0 \le f \le 1$ are allowed. 
 In contrast to the CMB, large scale structure data, which were initially used to rule out unclustered unified dark matter models, are far less constraining. Indeed, late universe data, including the large scale galaxy distribution, prefer models that are far from $\LCDM$. But these are in tension with the CMB data. 

\end{abstract}

\maketitle

\section{Introduction}
Shortly after it became clear that cosmological observations may require dark energy, as well as dark matter, 
to be added to the cosmic budget, 
the possibility that a single fluid may represent both constituents was proposed~\cite{AlterKame2001,Bento2002}. Such an option  is evidently appealing in view of its economy. It is also realisable, in principle, in the form of a fluid endowed with negative pressure $p$ that vanishes with increasing density $\rho$. Early on, when the universe is dense, 
the cosmological fluid would  behave as pressureless dark matter, and as the universe expands it  ultimately behaves as 
dark energy.
It would also constitute pressureless dark matter (DM) if it nonlinearly collapses and eventually clusters into dense halos, forming a CDM-like sector that may hierarchically merge and host galaxies as in the standard 'double dark' $\LCDM$ scenario~\cite{PrimackR,frenkwhite}.  

For such purposes. the particular case of the standard Chaplygin gas, with $p \propto - 1/\rho$, seemed moreover to be well motivated from the theoretical viewpoint~\cite{jackiw2000particle, gorini2005chaplygin}.  
It soon became clear however that the sound 
speed in such an expanding 
homogeneous fluid is a steep function of 
the scale factor, and that the associated pressure forces become large enough to prevent late structure formation in the linear regime~\cite{sand2004}. Indeed, this was found to be the case not only for the standard Chaplygin gas, but for any generalized model with $p \propto - 1/\rho^{\alpha}$, unless $\alpha \lesssim 10^{-5}$, in which case   
the cosmological model is indistinguishable from the standard $\Lambda$CDM, with an  additional parameter $\alpha$ that 
lacked a physical basis and required fine tuning. 
Chaplygin models also seemed viable when $\alpha > 1$, but in this case the sound speed may become superluminal.  Otherwise,  for standard adiabatic perturbations, the unified dark matter linear power spectrum displays violent oscillations that are entirely ruled out~\cite{gorini2008}. 

It was then pointed out that the linear perturbation power spectrum of the baryons, which remain essentially pressureless on the relevant  scales, is much closer to that observed~\cite{2003PhRvD..67j1301B, gorini2008}. 
As the matter power spectrum used  to constrain unified dark fluid parameters was based on the luminous matter, represented by the galaxy distribution, this may appear to go some way in 'saving' such models. 
However, this relative success 
actually highlights the basic problem with the assumptions leading to the original methodology used to rule out unified dark matter models~\cite{sand2004}. This  considered the evolution of 
late stage linear perturbations --- that is large scale perturbations that are still in the linear regime at low redshift  --- without tackling the question regarding the fate of smaller scale perturbations, which in the standard cosmological model should have already long formed filaments, pancakes or fully collapsed into dark matter halos. In fact only a tiny fraction of the dark matter is not in such structures  at late cosmic times in standard cosmology~\cite{WhiteClus2018}. 
And not taking this into account leaves open 
the question of the very 
possibility galaxy formation in a unified fluid based cosmological context; for, 
in the standard scenario, galaxies form inside dark matter halos, and in an analogous model they should form in  dark fluid condensations that play a similar role. Therefore, although more 
complicated unified dark fluid 
models, involving a fast transition between 
dark matter and dark energy behavior, 
can avoid some of the pathologies 
related to the unclustered Chaplygin gas~\cite{fasttrans2010, Fasttrans2011}, considering 
clustered models still seems mandatory 
from the physical point of view.

The relevant questions thus posed are: i) Do 
nonlinear structures form in the context of 
simple unified dark matter cosmologies? And ii) Would their formation  modify the various observational constraints that entirely rule out standard Chaplygin gas cosmologies 
and tightly limit the allowed values of $\alpha$ 
(e.g.,~\cite{sand2004, gorini2008, park2010, 2013PhRvD..87h3503W})?

Under the assumption that the nonlinear 
clustering of the unified fluid 
 is indeed possible, it was
shown that the associated background cosmological evolution becomes viable, 
as long as the clustering 
is efficient; such that $\gtrsim 90 \%$  of the  fluid is 
transformed into nonlinear structures that form a pressureless DM-like component~\cite{2014PhRvD..89j3004A, ZantChap2022}. An analysis based on a simple spherical collapse model, with pressure gradients included,  furthermore showed 
that the transformation of smooth Chaplygin gas into self gravitating structures  is indeed possible, even if the perturbations are (Jeans) stable in the linear regime
(and are thus expected to oscillate and damp rather than grow). 
The key here being that the magnitude of the pressure in a unified dark sector gas decreases with density, as opposed to standard gases where the reverse is the case (and where linear stability generally implies nonlinear stability). This leads to what one may term a 'nonlinear Jeans scale', which interestingly turned out to be comparable to that of the smallest structures that may host galaxies; and  may therefore have consequences for 'small scale problems' plaguing galaxy formation in the context of standard cosmology, such as the apparent overabundance of small halos relative to observed dwarf galaxies count~\cite{delpopolo, bullock, SalucciAAR19}.  

An additional appealing aspect that transpired from considering the clustered unified fluid cosmologies
regards the associated value of the Hubble constant. 
Although the unclustered sector's contribution 
to the cosmic budget becomes similar to that of a 
cosmological constant as the universe 
expands, the transition to such behavior was found to involve an effective phantom regime  (in the sense of the effective dark energy density contribution growing with time)~\cite{ZantChap2022}. 
This allows for higher values of the 
Hubble constant while keeping the angular diameter distance to the cosmic microwave background (CMB) fixed. Rendering CMB estimates more consistent with locally measured ones.  

With this context in mind, we here consider a 
larger set of observational tests to apply to clustered Chaplygin gas models, as prototypical examples of simple unified dark sector cosmologies with a DM-like component. We thus include background evolution data, as well as full large scale structure (LSS) and  CMB data, and their cross-correlation. In this way, we revisit the viability of unified dark matter models, while taking into account the possibility, and necessity,  of the clustering of the fluid into nonlinear structures.

The basic feature of the scenario studied, as presented in the next section, is that it effectively splits the unified fluid into two sectors, interacting only through gravity.
In Section~\ref{sec:background}, 
we study the background evolution of such models, and in Section~\ref{sec:pertb} we present the linear perturbation equations for the two sectors. These are used to infer the resulting  CMB and LSS matter power spectra. 
Along with outlining the general properties of the models studied,
Section~\ref{sec:cluschap} also presents a 
discussion of their basic  
properties and predictions 
with respect to the data, focusing on the standard Chaplygin gas. 
To clarify the essential role of 
secondary anisotropies in the CMB, we 
keep the distance to last scattering, and the parameters determining its spectrum there, fixed. Section~\ref{sec:combconstr}  presents a detailed 
statistical analysis of the models. The various datasets employed and the methods used  there are listed in Section~\ref{sec:datasets}. 
{Appendix~\ref{app:Table}  summarizes the corresponding 
estimates of the main cosmological parameters.
We present our conclusions in Section~\ref{sec:Conc}.}

\section{Clustered Chaplygin gas cosmology}
\label{sec:cluschap}

\subsection{Background evolution}
\label{sec:background}

The generalized Chaplygin gas equation of state, relating the pressure and density, 
is given (with $c=1$) by
\begin{equation}
    p_{\rm Ch } = - \frac{A}{\rho_{\rm Ch}^\alpha}. 
\label{eq:state}
\end{equation}
When $A$ and $\alpha$ are positive this represents a fluid endowed with a negative pressure that decreases in magnitude with increasing density. 
It thus conforms to the basic requirements for a unified dark fluid. 
In a cosmological context, a homogeneous Chaplygin gas obeys the conservation equation 
\begin{equation} 
\dot{\rho}_{\rm Ch} + 3 H (\rho_{\rm Ch} + p_{\rm Ch}) = 0,  
\label{eq:cons}
\end{equation}
with $H$ the Hubble parameter. This leads to a background density evolution given by 
\begin{equation} 
\rho_{\rm Ch} (t) = \bigg[ A + \frac{B}{a^{3(1+\alpha)}}\bigg]^{\frac{1}{1+\alpha}}, 
\label{eq:density}
\end{equation}
which describes a universe that is matter dominated ($\rho \sim 1/a^3$) as 
$a \rightarrow 0$, but tends towards  constant density as $a \rightarrow \infty$. In asymptotic terms, therefore, it has the same 
background evolution as the standard model. 
More quantitatively, at early times,
the fluid behaves as  dark matter with density $B^{1/1+\alpha}/a^3$. This relation is indeed valid to precision of order $a_{\rm lss}^3$, with $a_{\rm lss} \sim 10^{-3}$,  at CMB last scattering. Thus, in order to  reproduce the CMB peaks as in the standard model, one needs to fix $B$ such that the effective dark matter density at last scattering is  $\rho_0 \Omega_c/a_{\rm lss}^3$, 
where $\rho_0$ is total current density (at $a=1$). 
Furthermore, currently  one has
 $A + B = \rho_{\rm Ch}^{1+\alpha} = \rho_0^{1+\alpha} \Omega_{\rm Ch}^{1+\alpha}$.  
To match these requirements, in a critical universe, 
one can thus define  
\begin{equation} 
A = \rho_0^{1+\alpha} \left(\Omega_{\rm Ch}^{1+\alpha} - \Omega_c^{1+\alpha} \right), \:\:\:
B = \rho_0^{1+\alpha} \Omega_c^{1+\alpha}, 
\label{eq:ABH}
\end{equation}
where 
\begin{equation}
\Omega_{\rm Ch} = 1- \Omega_b-\Omega_R,
\end{equation}
and 
$\Omega_b$ and $\Omega_R$,
are the the baryon and radiation fractions.

In a clustered Chaplygin fluid one assumes that a fraction $f$ of its energy  
density may nonlinearly grow into self gravitating structures, eventually forming 
halos akin to those in the standard cosmology~\cite{ZantChap2022}. This fraction would in general be an increasing function of time, but in the simplest scenario 
$f$ may be assumed to be a constant,  fixed with the development of the first structures at some 
$a_{\rm cl}  \ll 0.1$, 
with $1-f$ representing a component of Chaplygin fluid that 
remains nearly homogeneous. This latter component is the one that 
eventually acts as dark energy, while the clustering component may hierarchically collapse and cluster as in standard CDM-based cosmology. 

{Energy conservation now requires that
the total energy density be given by}~\footnote{The physical motivation for this form was discussed in~\cite{ZantChap2022} for the case of the standard 
Chaplygin gas (Section~IV.A).  As the first nonlinear structures form, the unified dark fluid splits into two sectors, one remaining nearly homogeneous 
and another made of the nonlinear structures.  
If this occurs sufficiently early, 
the split is essentially between two pressureless components, with densities $f B^{1/{ 1+ \alpha}} a^{-3}$ and
$(1-f) B^{1/{1+\alpha}} a^{-3}$. 
Subsequently, if $f$ is fixed,  the
quasi-homogeneous component's density evolves according 
to the first term in (\ref{Eq:rho_bac}), as this is simply the analogue of 
equation~(\ref{eq:density}). The clustered sector continues to evolve 
as a matter component, represented by the second term of~(\ref{Eq:rho_bac}),  
and thus $\rho_{\rm DS}$ satisfies a conservation
equation analogous to (\ref{eq:cons}); it is therefore an accurate representation 
of the total energy density of the dark sector, provided the bulk of the variation in $f$ occurs sufficiently early. 
This is further discussed and quantitatively
illustrated in connection to Fig.~\ref{fig:back_rho_varf}.}
\begin{eqnarray}    
    \rho_{\rm DS} &=& \rho_{\rm Ch} + \rho_{\rm DM}  \nonumber  \\
    &=& \left[A_{\rm cl} + (1-f)^{(1 + \alpha)}  B_{\rm cl} a^{-3(1 + \alpha)}  \right]^{\frac{1}{(1 + \alpha)}} \nonumber \\
    && + {\bf  B_{\rm cl}^{\frac{1}{(1 + \alpha)}}~f~a^{-3}},
    \label{Eq:rho_bac}
\end{eqnarray}
where the subscript DS refers to the
total dark sector density. 
The first term here represents the 
component that remains unclustered. 
We accordingly refer to it, as before, as 
$\rho_{\rm Ch}$. The second terms 
is identified with the  
pressureless dark sector that behaves as standard 
dark matter, 
and we refer to it with the label DM.
{The constants $A_{\rm cl}$ and $B_{\rm cl}$ must again be chosen so as to reproduce the asymptotic limits of the standard model.}  
We therefore repeat the same exercise above to fix these constants in this new context. 

As early on, before the formation of the first halos ($a \ll 0.1$), 
the gas is unclustered and behaves as a nearly homogeneous dark matter component, the constant $B$ should remain unchanged, as 
given by the second equality in~(\ref{eq:ABH}); thus, $B_{\rm cl} = B$. 
This is again necessary to maintain a dark matter component at last scattering that corresponds 
to a mass density $\rho_0 \Omega_c/a_{\rm lss}^3$.
The relation analogous to the first  of the aforementioned equalities, on the other hand, becomes
\begin{eqnarray}
A_{cl} &=& \rho_0^{(1 + \alpha)} \left(\Omega_{\rm Ch}^{(1 + \alpha)} - (1 - f)^{(1 + \alpha)} \Omega_{c}^{(1 + \alpha)}\right),
%
\label{eq:Aclus}
\end{eqnarray}
where $\Omega_{\rm Ch}$ now refers 
only to the unclustered component and is given by
\begin{equation}
    \Omega_{\rm Ch} = 1 - \left(\Omega_b + \Omega_R + f\Omega_{c}\right). 
\end{equation}

In general, the Chaplygin gas equation of state parameter $w_{\rm Ch}$ is given by
\begin{equation}
    w_{\rm Ch} \equiv \frac{p_{\rm Ch}}{\rho_{\rm Ch}} =  -\frac{A}{A + \frac{B}{a^{3(1+\alpha)}}}.
    \label{eq:eos}
\end{equation}
Although this tends to $-1$ and does not fall below that, due to the nature of its 
time variation the background evolution displays a phantom-like
behavior, in the sense  of the existence of a rising effective dark energy. 
This is associated with the advent of a 
dark energy component at later times, as 
$w \rightarrow -1$, starting from an initial near zero value. 

{For the case of the clustered fluid, an equation analogous to~(\ref{eq:eos})
applies to the component that remains nearly homogeneous, such 
that
\begin{equation}
  \omega_{\rm Ch} = -\frac{A_{\rm cl}}{A_{\rm cl}+\frac{(1-f)^{1+\alpha} B} {a^{3(1+\alpha)}}}   
  \label{eq:eosc}
\end{equation}
(where we have used the result $B_{\rm cl} =B$). 
As we will see explicitly below, 
the presence of this component entails the possibility  of keeping the angular diameter distance to the CMB constant while increasing $H_0$ and decreasing $f$.} 
The CMB spectrum at last scattering
may also remain invariant if all other 
cosmological parameters are kept fixed. 
To keep the effective dark matter density 
at last scattering fixed while increasing 
$H_0$, one can decrease the relative dark matter 
density $\Omega_c$, such  $\Omega_c h^2$
remains invariant ($h = H_0/ 100 {\rm km/s/Mpc}$). 

\subsubsection{The Hubble and equation of state parameters}
\label{sec:hubble}

\begin{figure}
    \centering
    \includegraphics[width=\linewidth]{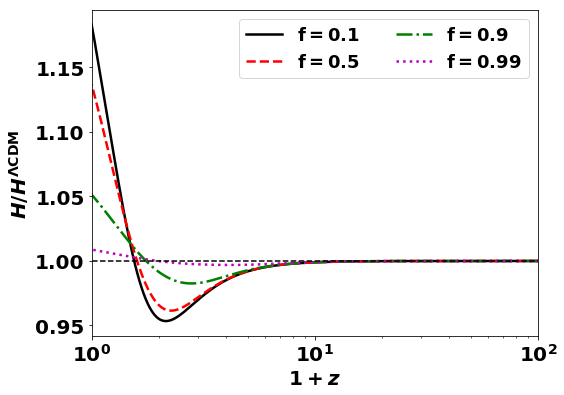}
    \caption{Evolution  of the Hubble parameter, relative to $\LCDM$ values, for different clustered unified fluid fraction $f$. The standard Chaplygin gas ($\alpha=1$) is assumed, and the transition to the clustered state  occurs at $a= 0.02$ ($z =49$). 
    As $f$ varies, all cosmological parameters are fixed at fiducial $\LCDM$ (Planck 2018~\cite{2020A&A...641A...6P}) 
    values, except for $\Omega_c$ 
    and $H_0$. These are chosen so as to keep the angular diameter distance to CMB last scattering, as well as the physical matter density there ($\propto \Omega_c H_0^2$), invariant.
    Note that $H_0$ increases as $f$ decreases.
     }
    \label{fig:Hubble_varf}
\end{figure}

\begin{figure}
    \centering
    \includegraphics[width=\linewidth]{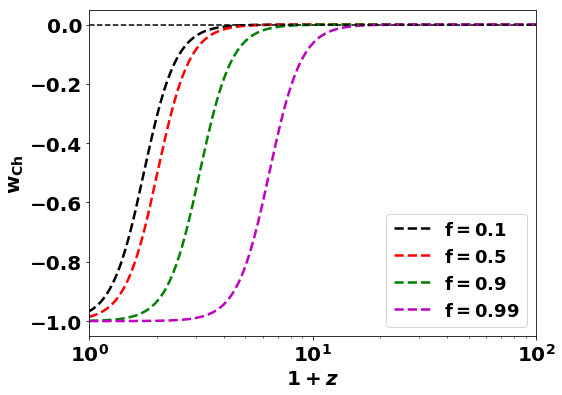}
    \caption{Equation of state parameter of the unclustered standard Chaplygin gas component for various fractions $f$ (for the same parameters as in  Fig.~\ref{fig:Hubble_varf}).}
    \label{fig:EoS_varf}
\end{figure}

The evolution of $H(z)$ for such models, relative to the standard scenario, is shown Fig.~\ref{fig:Hubble_varf}. All parameters are fixed to fiducial $\Lambda$CDM values, 
except for $H_0$ and $\Omega_c$, which are varied with the aforementioned considerations in  mind; 
to keep $\omega_c = \Omega_m h^2$ 
and the angular diameter distance to the CMB last scattering surface,
\begin{equation}
D_A (z_{\rm lss}) =  \frac{1}{1+z_{\rm lss}} \int_0^{z_{\rm lss}}  \frac{d z}{H (z)}, 
\label{eq:DA}
\end{equation}
fixed. 
Since the Hubble parameter is smaller than in $\Lambda$CDM for a significant fraction 
of late time evolution, it is possible to have for it 
to have higher as $z \rightarrow 0$ 
while keeping the integral in~(\ref{eq:DA}) invariant. 
{ This allows for higher values of $H_0$.}

The corresponding behavior of the equation of state parameter is 
illustrated in Fig.~\ref{fig:EoS_varf}. 
Note also that, as $f \rightarrow 1$ (more efficient clustering), 
the transition to a $\Lambda$-like behavior occurs earlier, when 
the dark energy component's contribution to the total energy budget is negligible. In this sense, the evolution of the clustered Chaplygin 
models tends to that of $\LCDM$ as $f \rightarrow 1$. 

\subsubsection{Density evolution}
\label{sec:dens}

\begin{figure*}
    \centering
    \includegraphics[width=0.49\linewidth]{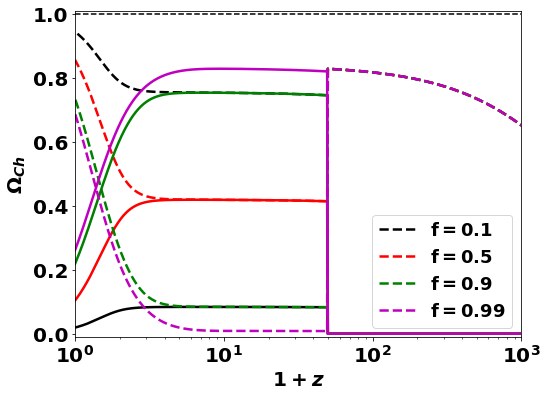}
    \includegraphics[width=0.49\linewidth]{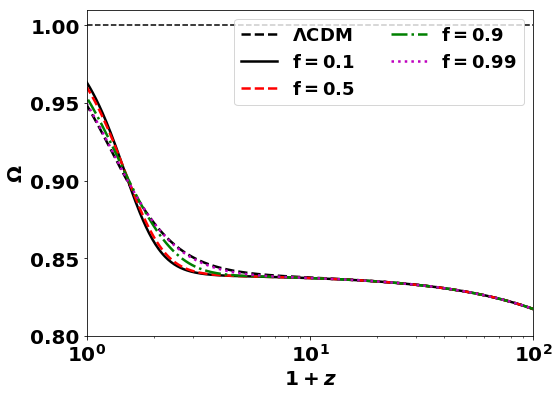}
    \caption{The left hand panel shows the evolution of the relative energy densities of unclustered
    Chaplygin gas (dashed lines) and clustered, pressureless DM-like component (solid lines),  for different  clustered fractions $f$ (and cosmological parameters  fixed as in Fig.~\ref{fig:Hubble_varf}). The right hand panel shows the two contributions added up. Note that although the clustering transition occurs relatively early on ($a= 0.02$, $z = 49$), significant differences in the curves only occur when $z$ is of order unity, 
    as up to that point both components behave as pressureless matter. Then the unclustered sector expands sufficiently so as to behave as dark energy. This latter transition occurs earlier as $f \rightarrow 1$. Note also that even for $f \rightarrow 1$, the unclustered sector eventually becomes dominant. This is analogous to late time $\Lambda$ domination in the standard model, despite its negligible early contribution. 
    } 
    \label{fig:back_rho_varf}
\end{figure*}

An important point to note regards the continuity 
of the total density, even when assuming an abrupt 
transition to a clustered system. 
As before, we assume a sharp transition and set it 
at $a = 0.02$ ($z = 49$), which approximately corresponds 
to the first nonlinear structures forming at the 
'nonlinear Jeans scale' inferred in~\cite{ZantChap2022} (about a comoving kpc). Nevertheless,  as can be seen from Fig.~\ref{fig:back_rho_varf}, 
the total density remains 
effectively continuous at the transition, 
regardless of the value of $f$. Indeed, measurable 
differences between the total densities of 
 models with quite different  fractions 
 of the components only occurs much later, 
when the unclustered component starts behaving 
as dark energy. 
Before that, both components effectively behave 
as dark matter in terms of its background density 
evolution, for this aspect they can be considered 
as one component.

In the following, we will invariably assume a sudden transition
that occurs at the aforementioned redshift ($z = 49$).
Physically speaking, this
is clearly an approximation; for, when 
the presence of a minimal scale is taken into account, 
the fraction of a clustered component 
in a matter dominated universe increases with time~\cite{{WhiteClus2018}}. 
But given the invariance of the density 
with $f$ until lower redshifts, a 
constant-$f$ transition 
that occurs at high redshift may not significantly 
affect the results, as long as the chosen corresponds to the 
eventual value expected at $z \gtrsim 1$. 
In our case, with its comoving kpc minimal scale,  the clustering fraction may be expected to be at least that 
of the 10~kev warm dark matter cosmology studied in~\cite{WhiteClus2018}
(which has an effective cutoff in the matter power spectrum at $k$ a few hundred ${\rm Mpc}^{-1}$). 
In that case, it was found that the unclustered fraction (in the terminology of that work, 
the 'one stream component'), corresponds to 
$1-f \approx 10^{-2}$ at low redshift. 

Below, we will furthermore see that the time dependence of the 
perturbed gravitational potentials, 
like the unperturbed background density, 
also only depend significantly on $f$ at late times. Taken together, 
this supports the use of the abrupt $f$-transition as an approximation. 

\subsection{Linear perturbations}
\label{sec:pertb}

We consider adiabatic perturbations of a perfect fluid 
in the Newtonian gauge of a flat Friedmann Robertson Walker metric,
\begin{equation}
    ds^2 = a^2 \left\{-(1 + 2\psi) dt^2 + (1 - 2\phi) dx^i dx_i \right\},
\end{equation}
where $\psi$ and $\phi$ are scalar potentials.  With no anisotropic stress, the linear perturbation equations are 
are~\citep{1995ApJ...455....7M}
\begin{eqnarray}
    \dot{\delta} &=& -(1 + w) \left(\theta - 3\dot{\phi}\right) - 3\left(c^2_{s} - w\right) {\cal H} \delta  \\
    \dot{\theta} &=& - (1 -3c^2_{s}) {\cal H}\theta + \frac{c^2_{s}k^2}{1 + w}\delta + k^2\psi, 
    \label{Eq:pert}
\end{eqnarray}
where $\delta = \delta \rho/\rho$ is the overdensity, and $\theta$ the velocity perturbation in the fluid. The derivatives are with respect to conformal time $d \tau = d t /a(t)$. 
The comoving Hubble function ${\cal H} = aH$, $w$ and $c_{s}$ are the equation of state parameters and the adiabatic sound speeds respectively. 

Before the clustering transition, the dark sector component 
density is given by 
Eq.~(\ref{eq:density}). 
After the transition, the density
of the unclustered and 
clustered components are represented by 
the first and second terms of Eq.~(\ref{Eq:rho_bac}), respectively.
For the clustered, DM-like component, 
we set $w = c^2_s = 0$.
For the  unclustered Chaplygin sector, the equation of state parameter $w_{\rm Ch} =p/\rho_{\rm Ch}$ is inferred { from Eq.~(\ref{eq:eosc}), with 
$A_{\rm cl}$} as given in (\ref{eq:Aclus}).
The adiabatic sound speed  is 
\begin{equation}
    c^2_s = \frac{\delta p_{\rm Ch}}{\delta \rho_{\rm Ch}} = \frac{\dot p_{\rm Ch}}{\dot \rho_{\rm Ch}} =  -\alpha w_{\rm Ch}.
\end{equation}
The relativistic Poisson equation is given by
\begin{equation}
    k^2 \psi = -4\pi G a^2 \rho \Delta, 
\end{equation}
where the total comoving density perturbations is  
\begin{equation}
    \rho_t \Delta = \sum \rho_i\left( \delta_i - 3H (1 + w_i) \theta_i\right).
\label{eq:poisson}
\end{equation}
Here the
index $i$ is understood to all components present (dark sector plus baryons and radiation).  
 
We have modified the publicly available Boltzmann code CLASS~\cite{2011arXiv1104.2932L},
adding the contributions of the clustered and unclustered 
Chaplygin gas to the background and linear perturbations, 
as outlined in the previous and current subsection. 
The contributions and interactions 
of other sectors (notably, the baryons and the radiation)
are included as in the standard code. The potentials in 
Eq.~(\ref{eq:poisson}) are in this way sourced by all components 
present. 

\begin{figure}
    \centering
    \includegraphics[width=\linewidth]{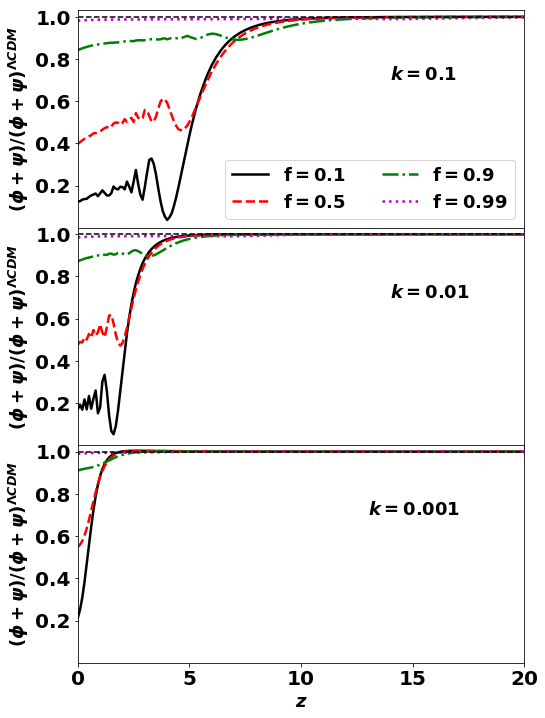}
    \caption{Newtonian potentials $\phi+\psi$ ratios, with respect to $\Lambda$CDM, as functions of redshift $z$ and the clustered fraction $f$. The potentials are calculated at scales $k = 0.1,0.01,0.001 \;\rm Mpc^{-1}$. The cosmological parameters are chosen the same way as in Fig.~\ref{fig:Hubble_varf}}. 
    \label{fig:Pot_varf}
\end{figure}

Fig.~\ref{fig:Pot_varf} shows the evolution for the  potentials in the same models 
discussed in the previous subsection (i.e., with parameters chosen so as to keep the distance to the CMB and spectrum at last scattering invariant). Again, as in the background evolution, there is little departure from the standard $\LCDM$ scenario, 
regardless of $f$, except when $z$ becomes of order unity or less. This is due to the late-time 
variation in $w_{\rm Ch}$, which is not present in $\Lambda$CDM.  
The discrepancy occurs later for smaller $k$, and so may be expected to have smaller imprint on the matter distribution.    
But the time variation in the potentials on such scales 
are probed by the integrated Sachs Wolfe effect (ISW) in the CMB~(e.g., \cite{SugiyamaHuSilk1996}).
This effect arises from the time variation in the potentials; it is given by~\cite{ISWREV2014PTEP}
\begin{equation}
\left(\frac{\delta T}{T}\right)_{\rm ISW} =
\int_{\tau_0}^{\tau_{\rm lss}} (\dot{\phi} + \dot{\psi)} d \tau. 
\end{equation}   
As in a critical matter dominated universe, with no dark energy, the potentials remain invariant on linear scales, the imprint of late time 
variations, probed by the ISW are a 
diagnostic of  the existence 
and properties 
of dark energy~\cite{ISWGalk1996PhRvL,LewisLnsingISW2022,ISW2024arXiv240106221S}.   
For the models at hand here, the strong 
time variations probed in Fig.~\ref{fig:Pot_varf} does  indeed lead to 
commensurably strong constraints on $f$, as we describe in what follows.

\subsubsection{The CMB, the ISW and lensing}
\label{sec:CMB_IWS}

\begin{figure}
    \centering
    \includegraphics[width=\linewidth]{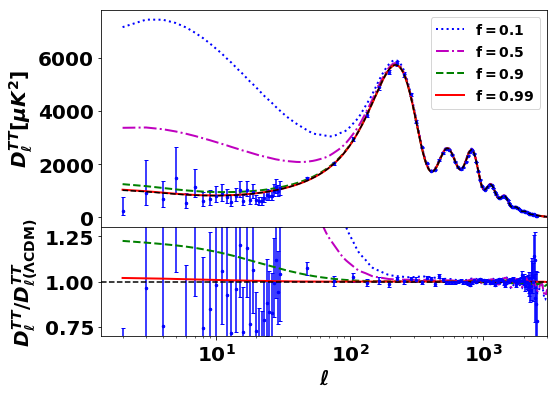}
    \caption{CMB Temperature anisotropy angular power spectrum, corresponding to the models considered in the previous figures.}
    \label{fig:ClTT_BestFit}
\end{figure}

\begin{figure}
    \centering
    \includegraphics[width=\linewidth]{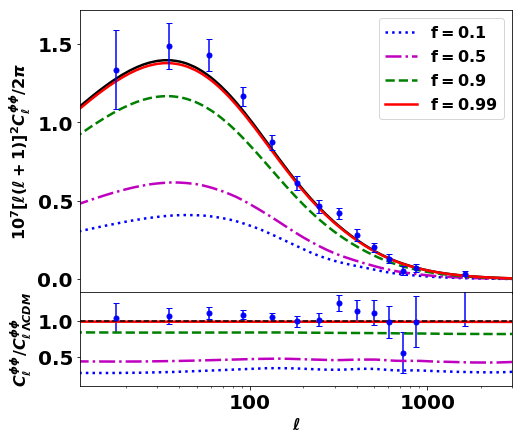}
    \caption{CMB lensing power spectrum for the same 
    models as in previous figures.}
    \label{fig:Clpp_BestFit}
\end{figure}

The CMB spectrum at last scattering is unaffected 
by changes in $f$, 
as long as the physical densities --- of baryons, radiation and of the effective 
dark matter component, represented at last scattering by the Chaplygin gas ---
remain fixed. Furthermore, as we have seen, the effect
of the background evolution on the location of the peaks (arising from variations in the angular diameter distance to last scattering), can be cancelled by increasing $H_0$. 

Even then, secondary anisotropies arising from late time 
perturbations remain. The ISW in particular turns out to constrain $f$ to remain quite close to unity, if the observed CMB spectrum is to be adequately reproduced. 
This is illustrated in Fig.~\ref{fig:ClTT_BestFit}, 
where we show the CMB power spectrum for the models that were the subject of the previous figures.  The spectrum turns out to be indeed effectively indistinguishable except for the enhancement 
on large scales (low $l$) due to the ISW, 
arising from the time variations in the gravitational 
potentials. These, as we saw, become quite large as $f$ departs significantly 
from unity 
(Fig.~\ref{fig:Pot_varf}).  
The effect of $f$ 
on the CMB lensing signal  is also constraining.
For $f \lesssim 0.9$ a clearly underestimated theoretical signal 
results (Fig.~\ref{fig:Clpp_BestFit}), as the effective 
DM fraction at late times decreases. 
However, due to the smaller number of data points 
the effect is actually 
less statistically significant than the ISW 
(as we will see in Section~\ref{sec:CMBstand}).

\subsubsection{Matter power spectrum}
\label{sec:MatterPS}

\begin{figure}
    \centering
    \includegraphics[width=\linewidth]{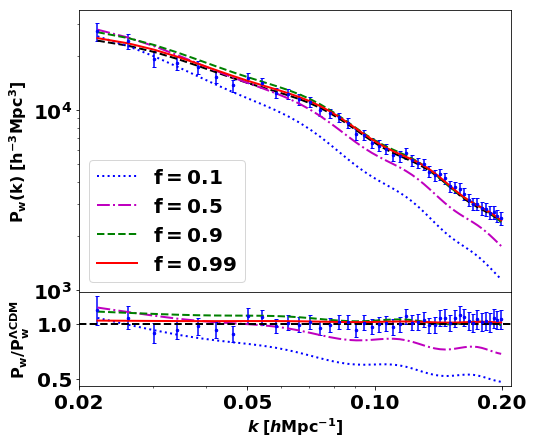}
    \caption{Matter power spectra in clustered standard Chaplygin gas cosmologies. The parameters are again fixed as in Fig.~\ref{fig:Hubble_varf} and following figures. The corresponding spectra are  compared to SDSS LRG DR7 data.}
    \label{fig:mPk_BestFit}
\end{figure}

It was originally the matter power spectrum that was used to 
tightly constrain, and effectively rule out, 
unified dark matter models~\cite{sand2004}. 
As mentioned in the introduction, the initial calculation 
ignored that, in any model where galaxy formation proceeds 
as in the standard scenario, the unified dark matter 
should collapse into halos inside which galaxies may form. 
The clustered Chaplygin gas model, entertained in the present work, takes 
this into account,  building upon previous results suggesting that 
nonlinear collapse of the unified dark fluid into self-gravitating structures is indeed possible. 
In this context, of the clustered Cahplygin gas, 
it turns out that the large scale structure signal is far less constraining than 
that of the CMB. 

We calculated the theoretical matter power spectra using the method 
described in Appendix~\ref{app:MPS}. 
The results are shown in 
Fig.~\ref{fig:mPk_BestFit}, where one immediately notes that the 
huge  oscillations present in 
the Chaplygin gas power spectrum (for $f=0$) are not there. 
Instead, the principal difference in the matter power spectrum of a clustered Chaplygin gas and baryon system
relative to $\Lambda$CDM
is the existence of 
damping on smaller scales. This arises from the strongly damped oscillations in the unclustered Chaplygin gas component at late times (see, e.g., Ref.~\cite{ZantChap2022} Fig.~5). 
However the ultimate effect of this damping is not very sensitive to $f$; 
and thus values of $f$ far smaller than allowed by CMB 
remain viable if large scale structure alone 
is included as a constraining dataset (as we will see more quantitatively in the next section).   

\section{Observational data sets and likelihood analysis}
\label{sec:datasets} 

Having probed the  essential features distinguishing  
clustered Chaplygin gas cosmologies, we now discuss in 
more detail the predictions of such models 
in relation to observations. 
Our goal is to conduct a generic likelihood analysis; 
evaluating  the probability of the models, 
given the 
constraints available with present data, both at the background evolution and linear perturbation levels. 

In doing so, 
we use the following datasets. 
\begin{itemize}
    \item 
    Background data: This includes distance modulus observations from SNIa sample Pantheon+ \citep{2022ApJ...938..110B}; the local measurement of the Hubble constant $H_0$, from the Hubble Space Telescope and the SH0ES Team\citep{2022ApJ...934L...7R}; and the prior on the baryon physical density $\omega_b = 0.0222 \pm 0.0005$, based on the measurement of D/H by \citep{2018ApJ...855..102C}, assuming standard big bang nucleosynthesis (BBN) and including modeling uncertainties. Hereafter we denote these datasets by "SNIa", "BBN", and "H0".
    
    \item 
    CMB data: We use the full temperature power spectrum of Planck 2018 data \citep{2020A&A...641A...6P}, adopting the baseline high-multipole likelihood. This includes high-$\ell$ multipole ($30 \leq \ell \leq 2508$) TT and high-$\ell$ ($30 \leq \ell \leq 1996$) EE and TE likelihoods. We also consider low-$\ell$ temperature and low-$\ell$ polarization likelihoods in the low-$\ell$ range ($2 \leq \ell \leq 29$). In addition, we use the CMB lensing-potential power spectrum. We denote the combined datasets by "CMB".
    
    \item 
    LSS data: 
    We use the matter power spectrum as measured from SDSS LRG DR7 \citep{2010MNRAS.404...60R}, which we denote by LSS. In addition, we use the CMB ISW-galaxy cross correlation power spectrum \citep{2008PhRvD..78d3519H,2018PhRvD..97f3506S}, denoted hereafter by "gISW". 
    
\end{itemize}

\subsection{Methodology}

As noted above, we modified the publicly available code CLASS \citep{2011arXiv1104.2932L} to incorporate the background evolution, as well as  
the perturbation equations described in Section~\ref{sec:pertb}.  
In this way, given $f$ and a set of cosmological parameters, we evaluate the background evolution of models and their CMB and LSS power spectra.  

We then perform full  likelihood analyses using the MontePython environment \citep{2013JCAP...02..001A} \footnote{https://baudren.github.io/montepython.html}. To scan the parameter space we use the Multinest mode \citep{2009MNRAS.398.1601F} and set the number of active points, $N$, to $1000$ and the sampling efficiency, $e$, is set to $0.8$. This  ensures a robust exploration of the posterior with an accurate evidence value.  To plot the posterior distributions and illustrate the confidence levels, we use GetDist python package\citep{2019arXiv191013970L}.

The parameter space is set to include the $\LCDM$ six vanilla parameters in addition to the clustered 
fraction parameter $f$ and Chaplygin equation of state index $\alpha$ (cf. Eq. \ref{eq:state}), namely: %
\begin{equation}\label{Prameter-space}
    \mathcal{P}\equiv\left\{\omega_{b},\, \omega_{c},\, H_0,\,
    \tau_{re},\, n_s,\, \log \left(10^{10} A_s\right), f, \alpha \right\}.
\end{equation}
The use of nested sampling as a likelihood scanner is motivated 
by the possibility of degenerate parameter space. In this case, it is preferable than posterior explorer methods (e.g. Metropolis-Hastings). The parameter space boundaries are set to be wide enough to ensure full exploration of possible degeneracy regions.  

The models, thus determined, are then confronted 
with various combinations of the datasets described above. 
The principal results are presented below. 
Appendix~\ref{app:Table} contains a table with the full 
set of inferred cosmological parameters and the
allowed values at the 
$68 \%$ confidence levels. 

\begin{figure}[H]
    \centering
    \includegraphics[width=\linewidth]{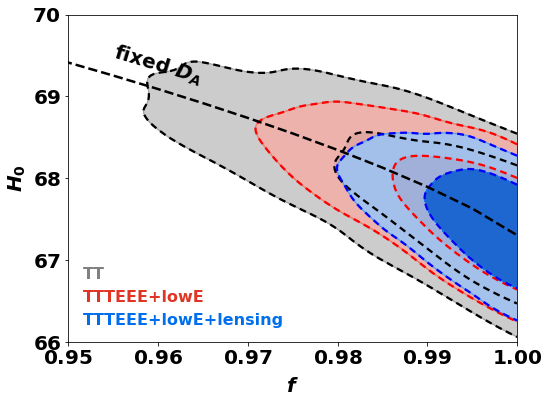}
    \caption{$f-H_0$ confidence contours 
    (at 68$\%$, 95$\%$ CL),
    emanating from Planck CMB datasets. The dashed black line is the relation between $f$ and $H_0$ given a fixed value of the distance $D_A$ (given by Eq. \ref{eq:DA}) to the CMB last scattering surface (i,e, with parameters corresponding to the set of models that are the subject of Fig.~\ref{fig:Hubble_varf} to 
    Fig.~\ref{fig:mPk_BestFit}).}
    \label{fig:fH0_CMB_confrgn}
\end{figure}

\begin{figure*}
    \includegraphics[width=0.4\linewidth]{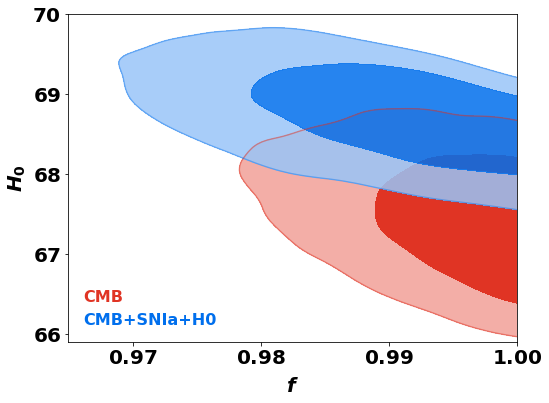}
    \includegraphics[width=0.59\linewidth]{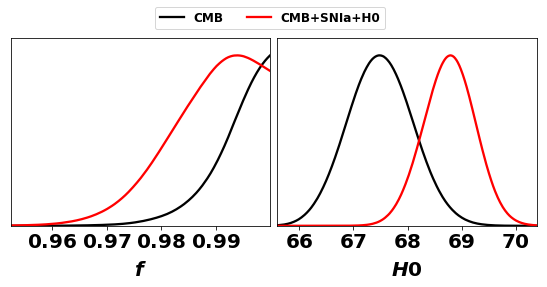}
    \caption{$f-H_0$ confidence contours 
    resulting from Planck 18 TTTEEE+lowE+lensing, PantheonPlus, and SH0ES (left panel), and corresponding probability density function (PDF)(right panel).}
    \label{fig:CMB_1d_fH0_PDF}
\end{figure*}

\section{Combined constraints and model likelihoods}
\label{sec:combconstr}

\subsection{CMB constraints on the standard gas}
\label{sec:CMBstand}

We begin by looking at the CMB constraints on the standard 
($\alpha =1$ in Eq.~\ref{eq:state}) Chaplygin gas. As expected from the discussion of the previous section, this puts tight limits on $f$. 
When all CMB data is included, the clustered fraction is 
constrained to be $f \gtrsim 0.99$ at the $68 \%$ confidence level, with preference for  $f \rightarrow 1$, which is effectively the $\Lambda$CDM model, which matches the mean of the PDF. This is illustrated in Fig.~\ref{fig:fH0_CMB_confrgn}, where we show the 
confidence contours constrained by CMB data~\footnote{In all the following figures
figures the confidence contours will also refer to $68 \%$ and $95 \%$ CL.}.
In the same figure, we also plot the constant angular 
diameter distance $D_A$ to the CMB line in terms of $f$ and $H_0$. It follows the contours on an upward path with decreasing $f$. This again reflects the relevance of such models to the $H_0$ tension. Even if, in their present form, 
the tension can only be mildly alleviated due to the tight constraints emanating from the ISW and lensing effects.  

To illustrate the point further, we explicitly 
include local background date, reflecting 
the familiar relatively high value of local measurments of $H_0$. The results 
are illustrated in Fig.~~\ref{fig:fH0_CMB_confrgn}. 
Now $f=0.99$ (rather than exact unity) becomes the  preferred value
(over $\LCDM$). Hubble constant values of up to $70 {\rm km/s/Mpc}$, and $f \simeq 0.97$, are allowed within the 
$95 \%$ confidence contour.

\begin{figure}[H]
    \centering   \includegraphics[width=\linewidth]{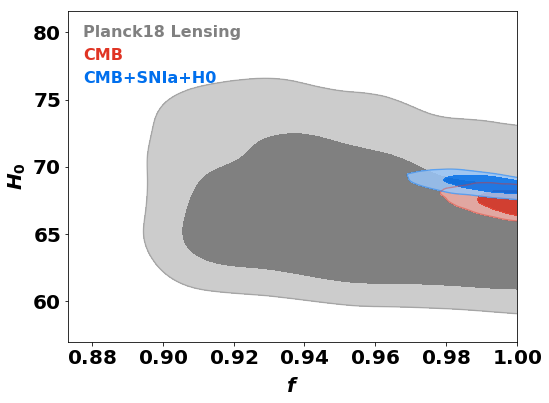}
    \caption{$f-H_0$ confidence contours, as constrained from Planck 18 lensing alone; Planck 18 TTTEEE+lowE+lensing; and when adding PantheonPlus, and SH0ES data.}
    \label{fig:CMB_conf_lens}
\end{figure}

As we will see below, these CMB constraints turn out to  provide the tightest limits on $f$ and $H_0$ and other cosmological parameters. Before we move 
to show this, we note, from Fig.~\ref{fig:fH0_CMB_confrgn} , that 
the added lensing constraints only 
moderately modify already strict limits on $f$ and $H_0$. Indeed, constraints from lensing 
are only significant in conjunction with limits imposed on the cosmological 
paramaters from other CMB data (Fig.~\ref{fig:CMB_conf_lens}).   

\subsection{Combined constraints on the standard gas}
\label{sec:CombStand}

\begin{figure*}[tp]
    \centering
    \includegraphics[width= 0.29\linewidth]{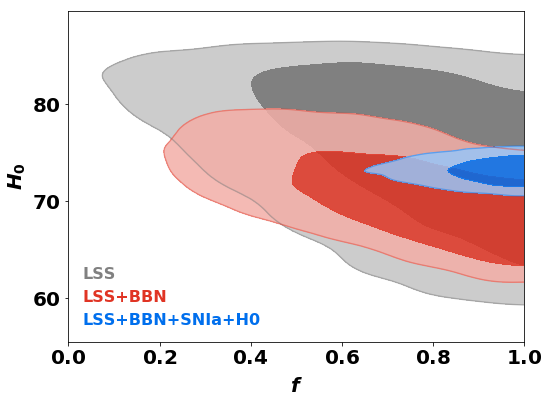}
    \includegraphics[width= 0.4\linewidth]{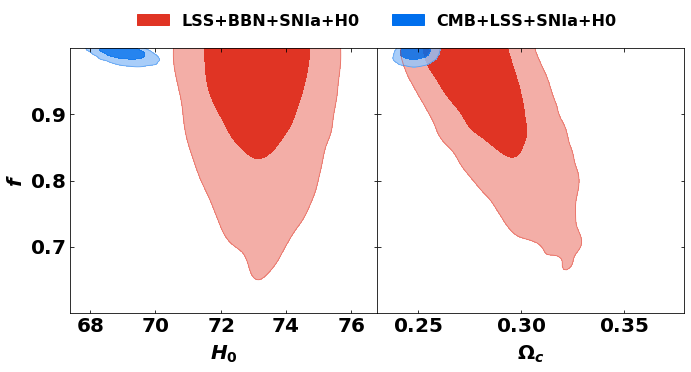}
    \includegraphics[width= 0.29\linewidth]{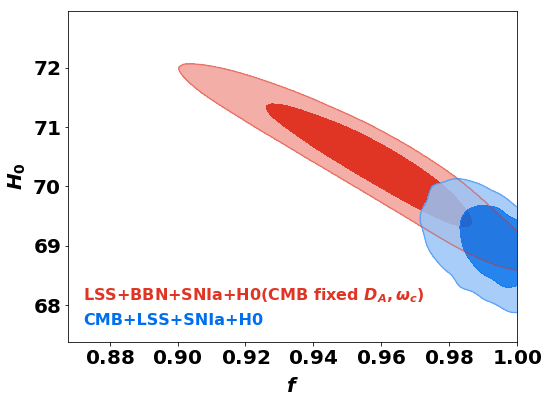}
    \caption{The left hand panel shows confidence contours using the LSS power spectrum from SDSS LRG DR7, alone and with the addition of PantheonPlus, BBN and SH0ES data. The middle panel illustrates the situation when the full CMB spectrum is included. It shows that though the combined late universe data (with BBN) favors large values of $H_0$, the resulting effective DM density at last scattering ($\propto \Omega_c H_0^2$) 
    is incompatible with CMB-inferred values. 
    The right hand panel shows that the constraints are significantly relaxed if, instead of using the full CMB spectrum at $z=0$, one uses fiducial Planck~18 values for the cosmological parameters while requiring that the distance to last scattering and $\omega_c = \Omega_c h^2$ remain invariant (as in the models discussed in Section~\ref{sec:cluschap}).}  
    \label{fig:LSS_conb}
\end{figure*}

\begin{figure}[h]
    \centering
    \includegraphics[width=\linewidth]{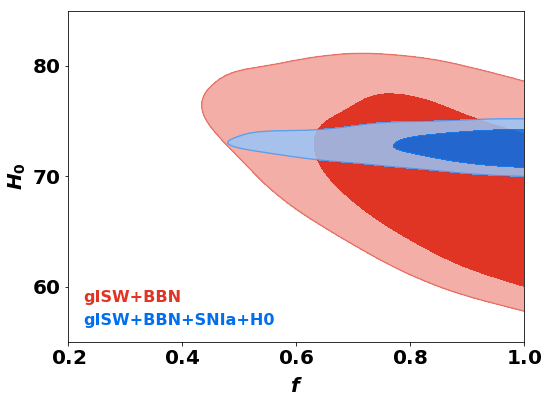}
    \caption{Confidence contours arising from galaxy ISW cross-correlation (using the methodology and datasets described in Appendix~\ref{app:gISW}).} 
    \label{fig:isw-gal}
\end{figure}

As already noted in Section~\ref{sec:MatterPS}, the power spectrum of the galaxy distribution, which was the original diagnostic used to rule out  unclustered unified dark matter cosmologies, is only weakly constraining the case of the clustered fluid considered here. 

This is reflected quantitatively in Fig.~\ref{fig:LSS_conb}. 
The left hand panel shows that 
even while including additional constraints, fixing the  
the baryon density and background evolution. values of $f \simeq 0.8$ are allowed at the $68 \%$ confidence level. These depart much further from unity  than allowed by the CMB constraints considered above. 
Within these limits, 
the large scale structure data, combined with local background evolution data, favours a large value 
for the Hubble parameter. But these limits are drastically decreased, along with $H_0$, when the CMB is included (middle panel).  
The reason is that in the models with 
large $H_0$ have 
physical matter density at last scattering which 
is too high to be compatible with the CMB. 
On the right hand panel 
we do as in in Section~\ref{sec:cluschap}; 
instead of including the full CMB spectrum 
we just add the constraint that the angular diameter distance to the surface of last scattering and the physical matter density 
there remain invariant. 
In this case, values of $f \simeq =0.9$ and $H_0 = 72~{\rm km/s/Mpc}$
are allowed within the 
$95 \%$ confidence contours. This suggests that models similar 
to the clustered Chaplygin gases, but with smaller ISW effect, may substantially alleviate the Hubble tension. 

Since the local galaxy distribution and  CMB photons, incoming 
from scattering, share the same gravitational potentials, an additional test of late time evolution is embodied in the galaxy distribution-ISW cross correlation~\cite{ISWGalk1996PhRvL}. 
We include this additional diagnostic in Fig.~\ref{fig:isw-gal}, where we calculated the cross correlation
as described in Appendix~\ref{app:gISW}. 
It again gives, at best, a weak constraint of $f \gtrsim 0.8$. But such tests may become stronger with ongoing and future LSS surveys. We note that
the gISW data, combined 
with late universe and BBN constraints, also favours a  high value of $H_0$. The 
contours are indeed similar to those inferred from LSS, BBN and late universe expansion in the previous figure. But they are likewise incompatible with CMB full power spectrum constraints.   

\begin{figure}[H]
    \centering
    \includegraphics[width=\linewidth]{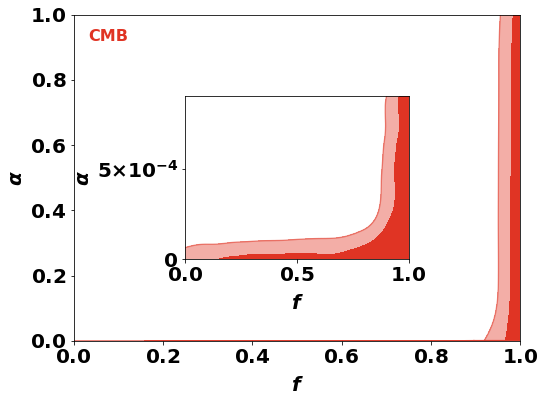}
    \caption{CMB constraints on the generalized Chaplygin gas, with equation of state given by (\ref{eq:state}) and $\alpha$ free to vary between zero and unity.}
    \label{fig:f_vs_alpha_contour_full}
\end{figure}

\subsection{Constraints on the generalized gas}

\begin{figure*}
    \centering
    \includegraphics[width=0.48\linewidth]{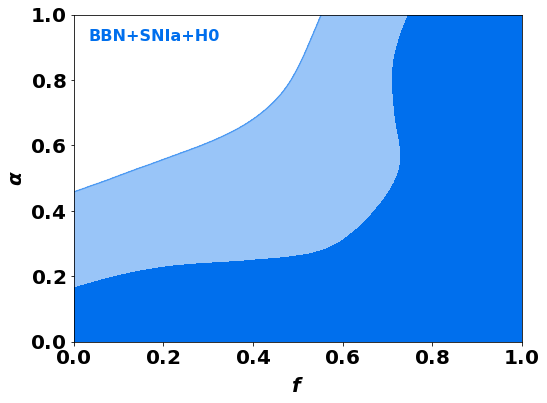}
    \includegraphics[width=0.48\linewidth]{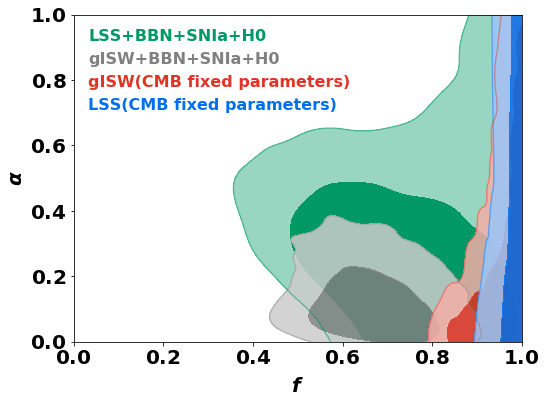}
    \caption{Left panel: $f-\alpha$ confidence contours from SNIa PantheonPlus data with prior constraints from BBN and SH0ES. Right: Adding the LSS galaxy power spectrum and galaxy ISW correlations. CMB fixed parameters refers to fixing all cosmological parameters to the CMB-inferred values, for a given $\alpha$ and $f$.}
    \label{fig:falpha_comb}
\end{figure*}

We now turn to the more general case of a unified dark fluid  described by the equation of state~(\ref{eq:state}) with $0 \le \alpha \le 1 $. 
We show the corresponding likelihood 
constraints from the CMB  in Fig.~\ref{fig:f_vs_alpha_contour_full}.
The tight limits on the minimal value of $f$ that we saw 
in the case of the standard gas 
turn out to persist for almost all $\alpha$, 
except for extremely small values, when the model also tends 
to $\LCDM$. For such small  $\alpha < 10^{-4}$, 
all values of $f$ are allowed. As has long been pointed out (e.g.~\cite{sand2004}),
this limit appears as an 
ad hoc representation 
of $\LCDM$. 
As it also requires much computational 
effort to zoom into the relevant region and obtain the corresponding 
confidence levels, we will not be considering it further
in what follows. 

As opposed to the CMB, 
the local large scale galaxy distribution tends to prefer values that are well removed from either $\LCDM$ limits. This is illustrated in 
Fig.~\ref{fig:falpha_comb}; when only background and BBN data are included as constraints, most values of both $f$ and $\alpha$ are allowed, with a preference for the standard model.   But adding  the spatial clustering of galaxies --- through the LSS power spectrum or the galaxy-ISW cross correlation ---  produces  tighter constraints. The distributions  are furthermore centered on values of $f \approx 0.7$. With corresponding $\alpha \approx 0.3$ for the LSS case and $\alpha \approx 0.1$ for ISW galaxy correlation. These correspond to cosmologies 
that are different from the standard model. 

\begin{figure}[h]
    \centering
    \includegraphics[width=\linewidth]{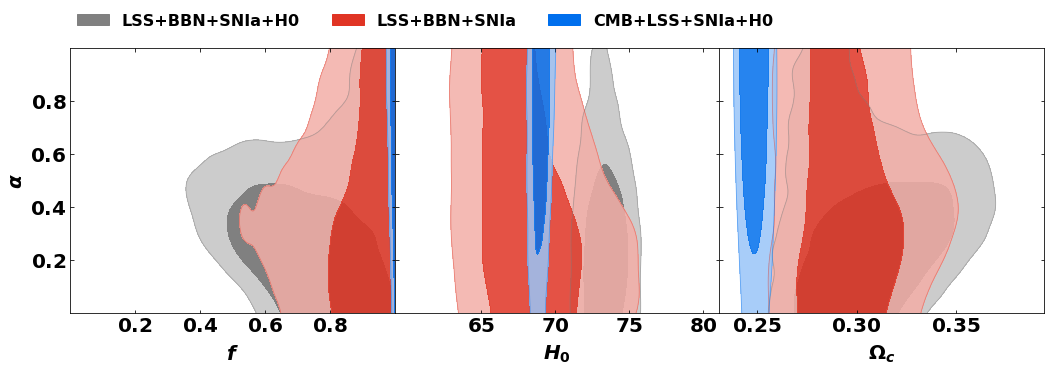}
    \caption{Confidence regions of generalized Chaplygin gas cosmologies from LSS data and combinations of BBN, SNIa, $H_0$, and CMB datasets.}
    \label{fig:falpha_All_LSS_data}
\end{figure}

\begin{figure*}
    \centering    
    \includegraphics[width=\linewidth]{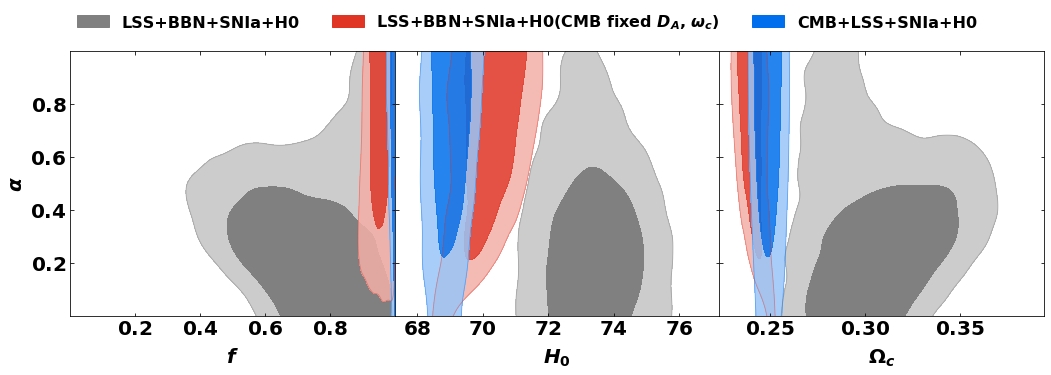}
    \caption{Constraints on generalized Chaplygin gas parameter $\alpha$, clustering fraction $f$, and effective dark matter fraction at last scattering $\Omega_c$.  Grey contours show constraints from LSS, BBN and late universe data. Blue contours include, in addition, the full CMB power spectrum. Red contours allow for values of $H_0$ and 
    $\Omega_c$ that keep the distance to CMB last scattering $D_A$, and $\omega_c = \Omega_c h^2$, invariant (as in the models of Section~\ref{sec:cluschap}), without   including the full CMB spectrum.       
    }
    \label{fig:falpha_combf}
\end{figure*}

As with the case of the standard Chaplyging gas considered above, the  
associated cosmological parameters are again not compatible with those inferred from the CMB.
We show this in the case of LSS data in Fig.~\ref{fig:falpha_All_LSS_data}.
But, once again, fixing the distance to the CMB and all CMB inferred physical 
densities, while not imposing the full CMB spectrum, leads to much 
looser constraints (Fig.~\ref{fig:falpha_combf}).
In this case, the combined constraints tend 
to again favor models that are different from $\LCDM$, 
and which come with higher Hubble constants. 
This  once more raises the possibility that models similar 
to those considered here but with significantly subdued ISW effect 
may be favored by cosmological data, particularly by the 
large scale galaxy distribution. At any rate, it may be a possibility worth testing with large amount of incoming data from ongoing and future galaxy surveys. 

{
\subsection{A note on the Hubble tension}
\label{sec:tension}

We finally comment on the consequences of the
models discussed  in this work on 
the conflict between the CMB-inferred and late expansion measurements of $H_0$ based on supernovae and data anchored to Cepheids, which appear to be in $5-\sigma$ tension with each other in the 
context of $\LCDM$~\cite{2022ApJ...934L...7R}. 
As discussed in
Section~\ref{sec:hubble}, when $f$ decreases away from unity, it is possible to keep the angular diameter distance to the CMB constant, while 
also keeping the spectrum at last scattering fixed and increasing $H_0$. This completely solves the $H_0$ problem, as stated above, if it were not for secondary anisotropies, particularly the ISW effect, which constrain $f$ to be very close to unity (cf. Section~\ref{sec:CMBstand}). 
In this case, the tension with SHOES data~\citep{2022ApJ...934L...7R} is only modestly  milder than in $\LCDM$: $4.3 \sigma$, for both 
standard ($\alpha = 1$ and generalized Chaplygin gases. 
There is no significant tension between the SHOES  value of $H_0$ and the 
values inferred from various datasets used here, when these datasets  do not include CMB input.     
The tension fully 
returns at the $4 \sigma$ level 
when the full CMB spectrum at $z=0$ (including secondary anisotropies) 
is included. 
Without the secondary anisotropies, with  parameters satisfying
only constraints from the CMB spectrum at last scattering, 
the tension is significantly reduced; 
for LSS power spectrum with CMB parameter combination, 
for example, the tensions with the SHOES value of $H_0$ are $3.2-\sigma$ and~$2.8-\sigma$, for the generalized and standard Chaplygin models, respectively. 
The remaining tension is due to the effect of fixing 
of $\omega_c$ from CMB data (as illustrated in Fig.~\ref{fig:falpha_combf}).  
The reduction in the Hubble tension when secondary anisotropies in the 
CMB are not included may
further motivate the exploration of models similar to those discussed here,
but with potentially smaller effects on the secondary anisotropies.}

\section{Conclusion}
\label{sec:Conc}

We have considered clustered unified dark fluid cosmologies, with emphasis on the standard
Chaplygin gas and using various observables to study the associated parameter space. 
In these models, a fraction $f$ of the unified fluid is assumed to be able to nonlinearly collapse into 
self gravitating systems, eventually forming structures akin to halos in standard CDM-based cosmology. This builds on a previous study, where it was shown that such collapse may be 
possible above a 'nonlinear Jeans scale' of the order of a kpc~\cite{ZantChap2022}. 
This is a necessary and sufficient condition for galaxy formation to proceed as in
standard cosmology; that is, inside self gravitating dark 
structures that may hierarchically cluster. 
We use, for simplicity, a constant fraction $f$ and discuss the  plausibility
of that assumption in Section~\ref{sec:dens}. 
The resulting models tend to the standard cosmology as $f \rightarrow 1$. Both in terms of background evolution, as shown in ~\cite{2014PhRvD..89j3004A, ZantChap2022} and confirmed here, as well as in terms of linear perturbations, as shown in Section~\ref{sec:pertb} of the present work (Fig.~\ref{fig:Pot_varf}). 

In Section~\ref{sec:hubble}, we  showed that the models considered here 
have the appealing property of allowing 
larger $H_0$, while keeping the distance to the CMB last 
scattering surface and the CMB temperature spectrum 
there invariant. 
Nevertheless, secondary anisotropies ---  
the Integrated Sachs Wolfe effect 
and, to a lesser extent, lensing --- 
cause the models to be highly constrained. In particular, the allowable values of the clustered fraction $f$ must be quite close to unity. {A further discussion of this effect is presented in Section~\ref{sec:tension}}.  

As opposed to the CMB, data from the large scale galaxy distribution, which were initially used to rule out unclustered unified dark matter systems~\cite{sand2004}, provide far looser constraints when clustered systems, which effectively include a DM-like component, are considered.  

To quantify these general trends, 
we confront the models with various cosmological observables, both in terms of background evolution 
and linear perturbations. 
The analysis, presented in Section~\ref{sec:combconstr},  
confirms the basic trends discussed above. Only models with extremely 
efficient clustering ($f \rightarrow 1$), and therefore quite close 
to standard cosmology, are allowed by the full CMB spectra.
And the CMB alone prefers values tending exactly to unity.
Nevertheless, when local universe 
background evolution data 
is included, a value of $f =0.99$ is preferred over systems coinciding with $\LCDM$.
The corresponding $H_0$ is also measurably larger than in the standard model; with values up to 
$H_0 \simeq 70 {\rm km/s/Mpc}$ (associated  with
$f\simeq0.97$) allowed at the $95 \%$ confidence 
level  (Fig.~\ref{fig:CMB_1d_fH0_PDF}).

The aforementioned trends are generally reproduced in the case of the generalized Chaplyging gas. Except when~$\alpha \rightarrow 0$ in Eq.~(\ref{eq:state}), 
in which case all values of $f$ are allowed 
(Fig.~\ref{fig:f_vs_alpha_contour_full}).  
Given the extra parameter $\alpha$ that requires fine tuning, 
this may be considered a rather ad hoc representation of $\LCDM$.
The limit of $f \rightarrow 1$, on the other hand,  has a 
definite physical interpretation, in terms of extremely efficient 
clustering, which may in principle be testable. 

The fraction of 
dark matter  that is comprised in nonlinear objects
has been studied in standard cosmologies~\cite{WhiteClus2018}.
This fraction may be taken to correspond to our splitting in terms 
of the parameter $f$, as clustering into nonlinear objects 
would have occurred during matter domination, when the unified fluid effectively
acted as a matter component (cf. Section~\ref{sec:dens}). 
Although the CMB-inferred constraints on $f$ are quite tight, 
they may not be incompatible  with the corresponding 
fraction of dark matter in nonlinear objects in warm dark matter 
cosmologies with 10~kev thermal 
particles. These come with a streaming length 
that is moderately larger than the nonlinear Jeans scale associated with standard
Chaplygin gas that was found in~\cite{ZantChap2022}. For WIMPs, the 'one stream' 
fraction (corresponding to our $1-f$), composed of dark matter that is not included in
nonlinear objects, is estimated to be smaller still (of order $10^{-3}$ at low redshift~\cite{WhiteClus2018}).  

In the case of the generalized gas, the combined late-universe data favor models that are quite different from $\LCDM$ (Fig.~\ref{fig:falpha_comb}). 
This is a potentially interesting phenomenon that may be 
worth further testing with upcoming large scale structure and background evolution 
data. Although the inferred cosmological parameters are 
incompatible with the full CMB power spectrum, 
omitting the effect of secondary anisotropies
again leads to milder constraints, while maintaining a preference for 
models that significantly depart from $\LCDM$. 
This may  motivate a search for clustered unified dark fluid cosmologies 
that retain the appealing  
aspects of a single physical dark sector and larger $H_0$,   
but are associated with smaller integrated Sachs Wolfe effect. 

\begin{table*}
    \begin{adjustbox}{width=\textwidth,center}
    \centering
    \setlength\extrarowheight{2mm}
    \begin{tabular}{l|c|c|c|c|c|c|c|c}
    \hline\hline
      Dataset  &$f$ & $\alpha$ & $H_0$ & $\omega_{c}$ &$10^{-2}\omega_{b}$ & $\ln10^{10} A_{s}$ &  $n_{s }$ & $\tau{}_{\rm reio }$\\
    \hline
    \multicolumn{9}{c}{$\alpha = 1.0$ case} \\
    \hline
    SNIa+BBN+H0*&$0.759,\;> 0.710$&$-$&$73.79\pm 0.97$&$0.189^{+0.022}_{-0.028}$&$2.270\pm 0.037$&$-$&$-$&$-$\\
    Planck 18 Lensing &$0.950^{+0.037}_{-0.028}$&$-$&$67.0^{+3.6}_{-5.3}$&$0.1198^{+0.0020}_{-0.0026}$&$2.238^{+0.028}_{-0.024}$&$3.048\pm 0.023$&$0.9642^{+0.0067}_{-0.0080}$&$0.055\pm 0.012$\\
    CMB&$0.994,\;> 0.993$&$-$&$67.50\pm 0.54$&$0.1206\pm 0.0012$&$2.232\pm 0.014$&$3.049\pm 0.014$&$0.9639\pm 0.0040$&$0.0556\pm 0.0072$\\
    CMB+SNIa+H0&$0.989,\;> 0.986$&$-$&$68.77\pm 0.43$&$0.11846^{+0.00092}_{-0.0012}$&$2.253^{+0.015}_{-0.012}$&$3.059^{+0.017}_{-0.014}$&$0.9691^{+0.0042}_{-0.0035}$&$0.0623\pm 0.0068$\\
    LSS+BBN*&$0.731,\;> 0.657$&$-$&$70.5^{+3.6}_{-4.5}$&$0.145^{+0.016}_{-0.025}$&$2.272\pm 0.037$&$-$&$-$&$-$\\
    LSS+BBN+SNIa+H0*&$0.908,\;> 0.892$&$-$&$73.13\pm 0.93$&$0.1528^{+0.0086}_{-0.012}$&$2.280\pm 0.038$&$-$&$-$&$-$\\
    gISW+BBN*&$0.808,\;> 0.758$&$-$&$70.4\pm 4.8$&$0.126^{+0.012}_{-0.018}$&$2.258^{+0.036}_{-0.024}$&$-$&$-$&$-$\\
    gISW+BBN+SNIa+H0*&$0.876,\;> 0.860$&$-$&$72.69\pm 0.92$&$0.1360^{+0.0084}_{-0.019}$&$2.263^{+0.033}_{-0.021}$&$-$&$-$&$-$\\
    \hline
    \multicolumn{9}{c}{free $\alpha$ case} \\
    \hline
    SNIa+BBN+H0*&$0.604,\;> 0.470$&$0.419,\;< 0.569$&$73.78\pm 0.96$&$0.178^{+0.017}_{-0.023}$&$2.270\pm 0.037$&$-$&$-$&$-$\\
    LSS+BBN+SNIa*&$0.866,\;> 0.841$&$0.439,\;< 0.580$&$68.3^{+2.6}_{-3.4}$&$0.139^{+0.012}_{-0.020}$&$2.270\pm 0.037$&$-$&$-$&$-$\\
    LSS+BBN+SNIa+H0*&$0.73^{+0.21}_{-0.12}$&$0.308,\;< 0.374$&$73.39\pm 0.96$&$0.166^{+0.014}_{-0.017}$&$2.275\pm 0.037$&$-$&$-$&$-$\\
    gISW+BBN+SNIa+H0*&$0.665\pm 0.088$&$0.127,< 0.149$&$73.60\pm 0.97$&$0.160\pm 0.011$&$2.258^{+0.035}_{-0.024}$&$-$&$-$&$-$\\
    CMB+LSS+SNIa+H0&$0.992,\;> 0.990$&$0.623,\;> 0.510$&$68.99\pm 0.42$&$0.11855^{+0.00092}_{-0.0012}$&$2.251\pm 0.012$&$3.053\pm 0.014$&$0.9681\pm 0.0036$&$0.0600\pm 0.0070$\\
    LSS+BBN+SNIa+H0*{$\dag$} &$0.950^{+0.034}_{-0.021}$&$0.645,\;> 0.535$&$70.18\pm 0.66$&$-$&$-$&$-$&$-$&$-$\\
    \hline\hline
    \end{tabular}
    \end{adjustbox}
    \caption{Peaks of PDF (mean) and $68\%$ confidence level constraints on main cosmological parameters from different data sets discussed in the main text. In all datasets with *, the parameters $\tau{}_{\rm reio}$, $\ln 10^{10} A_{s}$ 
    and $n_{s}$ are fixed and set to values $\tau{}_{\rm reio} = 0.0544$,  $ln10^{10}A_{s} = 3.044$, and $n_{s} = 0.9649$ according to $\Lambda$CDM Planck~18 results \citep{2020A&A...641A...6P}. The data set with $\dag$ corresponds to the case where $D_A,\; \omega_{cdm}$, and $\omega_{b}$ are fixed.}
    \label{tab:param_1sgma_table}   
\end{table*}

\begin{acknowledgments}
We thank the referee for 
a careful reading and constructive comments and suggestions, 
and Roy Maartens for commenting on an earlier version of the manuscript. 
This paper is based on work supported by the Science and Technology and Innovation Funding Authority (STDF) under grant number {48289}. We would like to thank Waleed El Hanafy for the useful discussions.  
\end{acknowledgments}

\appendix
\section{Estimates of main cosmological parameters}
\label{app:Table}

In Table~\ref{tab:param_1sgma_table} we present the estimates of the main 
cosmological parameters (mean values of the PDF and
$68 \%$ confidence levels), for the various 
constraining datasets used. 
The corresponding likelihood 
contours are presented in Section~\ref{sec:combconstr}.  

\section{Matter power spectrum}
\label{app:MPS}

To compare the theoretical matter power spectrum 
with data, we evaluate the constrained convolved linear power spectrum, defined as 
\begin{equation}
    P_w(k_i, p) = \sum_n W(k_i, k_n) P(k_n, p) - W(k_i, 0).
\end{equation}
Here $W$ is a window function that relates the theoretical power spectrum, for cosmological parameters $p$ and evaluated at wavenumbers $k_n$,
to the central wavenumbers of the observed bandpowers $k_i$ \citep{2005MNRAS.362..505C}. The window functions were calculated as described in \citep{2001MNRAS.327.1297P, 2007ApJ...657..645P}. 
In Fig.~\ref{fig:mPk_BestFit} the comparison is made with the 
measured SDSS LRG DR7 power spectrum. This  has 45 band power measurements at comoving scales $k = 0.02 - 0.2 h/{\rm Mpc}$. The corresponding band-power window functions are measured along with the inverse covariance matrix between the measurements.

\section{Galaxy ISW cross-correlation}
\label{app:gISW}

In this appendix we briefly outline the method used to calculate
the galaxy CMB-ISW cross correlation. 
The cross-correlation angular power spectrum between CMB temperature anisotropy (T) and galaxies, as tracers of matter distribution (G), is given by~\citep{2018PhRvD..97f3506S}
\begin{equation}
    C^{TG}_{\ell} = \frac{3\Omega_m H^2_0}{(\ell +\frac{1}{2})^2} \int  b(z) \frac{dN}{dz} a H \frac{d}{dz} \left(\frac{P\left({\mathcal K},z \right)}{a^2}\right) dz. 
\end{equation}
Here, $b$ is the galaxy bias factor, $dN/dz$ the redshift distribution, $P({\mathcal K}, z)$ is the matter power spectrum at ${\mathcal K} = (\ell+1/2)/r(z)$, and $r(z)$ is the comoving radial distance. 

The CMB maps we used are from the Planck 15 data release. For the galaxy sample, the following catalogues of extragalactic sources are used: the 2MASS Photometric Redshift catalogue (2MPZ); the WISE$\times$SuperCOSMOS photo-z catalogue; the Sloan Digital Sky Survey Data Release 12 (SDSS-DR12) photo-z sample; 
a catalogue of photometric 
quasars (QSOs), compiled by \citep{2009ApJS..180...67R} from the SDSS
DR6 dataset (SDSS DR6 QSO); as well as the NRAO VLA Sky Survey (NVSS) catalog of radio sources.


\bibliographystyle{apsrev4-2}
\bibliography{chap_II}

\end{document}